\documentclass[twocolumn]{aastex6}

\newcommand\about  {\hbox{$\sim$}}
\renewcommand\deg  {\hbox{$^\circ$}}
\newcommand{\E}[1]{\hbox{$10^{#1}$}}
\newcommand\x      {\hbox{$\times$}}

\newcommand\Mo     {\hbox{$M_{\odot}$}}
\newcommand\kms    {\hbox{km\,s$^{-1}$}}
\newcommand\cc     {\hbox{cm$^{-3}$}}
\newcommand\cs     {\hbox{cm$^{-2}$}}
\newcommand\mone   {\hbox{$^{-1}$}}
\newcommand\pv     {\hbox{$p$--$v$}}

\begin{document}

\title{High-Velocity Bipolar Molecular Emission from an AGN Torus}

\author{Jack F. Gallimore\altaffilmark{1}, Moshe Elitzur\altaffilmark{2,3},
         Roberto Maiolino\altaffilmark{4,5}, Alessandro Marconi\altaffilmark{6,7},
         Christopher P. O'Dea\altaffilmark{8}, Dieter Lutz\altaffilmark{9},
         Stefi A. Baum\altaffilmark{8},
         Robert Nikutta\altaffilmark{10},
         C. M. V. Impellizzeri\altaffilmark{11},
         Richard Davies\altaffilmark{9},
         Amy E. Kimball\altaffilmark{12},
         and Eleonora Sani\altaffilmark{13}
         }

 \altaffiltext{1}{Dept. of Physics \& Astronomy, Bucknell University, Lewisburg,
     PA 17837; jgallimo@bucknell.edu}
 \altaffiltext{2}{Astronomy Dept., University of California, Berkeley,
     CA 94720}
 \altaffiltext{3}{Physics \& Astronomy, University of Kentucky, Lexington,
     KY 40506}
 \altaffiltext{4}{Cavendish Laboratory, University of Cambridge, 19 J. J. Thomson Ave, Cambridge CB3 0HE, UK}
 \altaffiltext{5}{Kavli Institute for Cosmology, University of Cambridge, Madingley Road, Cambridge CB3 0HA, UK}
 \altaffiltext{6}{Dipartimento di Fisica e Astronomia, Universit\`a di Firenze, via G. Sansone 1, 50019, Sesto Fiorentino (Firenze), Italy}
 \altaffiltext{7}{INAF-Osservatorio Astrofisico di Arcetri, Largo E. Fermi 5, 50125, Firenze, Italy}
 \altaffiltext{8}{University of Manitoba, Department of Physics and Astronomy, Winnipeg, MB R3T 2N2, Canada}
 \altaffiltext{9}{Max Planck Institute for Extraterrestrial Physics, Giessenbachstrasse 1, D-85748 Garching, Germany}
 \altaffiltext{10}{Instituto de Astrof\'isica, Facultad de F\'isica, Pontificia Universidad
Cat\'olica de Chile, 306, Santiago 22, Chile}
 \altaffiltext{11}{Joint ALMA Office, Alsonso de Cordova 3107, Vitacura, Santiago, Chile}
\altaffiltext{12}{National Radio Astronomy Observatory, 1003 Lopezville Rd.,
Socorro, NM  87801, USA}
\altaffiltext{13}{European Southern Observatory, Alonso de Cordova 3107, Vitacura, Santiago, Chile}

\begin{abstract}

We have detected in ALMA observations CO $J = 6\to5$ emission from the
nucleus of the Seyfert galaxy NGC~1068.  The low-velocity (up to $\pm$70
\kms\ relative to systemic) CO emission resolves into a 12\x7 pc structure,
roughly aligned with the nuclear radio source. Higher-velocity emission (up
to $\pm 400$~\kms) is consistent with a bipolar outflow in a direction nearly
perpendicular ($\simeq 80\deg$) to the nuclear disk. The position-velocity
diagram shows that in addition to the outflow, the velocity field may also
contain rotation about the disk axis. These observations provide compelling
evidence in support of the disk-wind scenario for the AGN obscuring torus.

\end{abstract}


\section{Introduction} \label{sec:intro}

The great diversity of AGN classes has been explained by a single unified
scheme. The nuclear activity is powered by accretion onto a super\-massive
(\about\E6--\E{10} \Mo) black hole. The active region is surrounded by an
obscuring, dusty toroidal structure so that sources viewed face-on are
recognized as type 1 AGN, and those observed edge-on are type 2
\citep[e.g.,][]{Ski93, Urry95}. However, the origin and nature of the obscuring
torus remain far from understood, with proposed models broadly divided into two
main classes. In one, the torus is a stationary doughnut-like structure
\citep{Krolik88}, and its large dimensions ($\ga$ 100 pc) place it well outside
the black hole (BH) radius of gravitational influence \citep{PK93, Granato94}.
In the other, a dynamic picture is employed: the torus is made of gas processed
by the accretion disk and expelled outwards in a disk wind; the outflow inner
part is atomic and ionized, giving rise to broad line emission, its outer
regions are molecular and dusty, resulting in a compact (well within the BH
gravitational potential), axi\-symmetric obscuring structure that mimics a
hydrostatic toroidal distribution \citep[][and references
therein]{Elitzur_Shlosman}.

%%%%%%%%%%%%%%%%%%%%%%%%%%%%%%%%%%%%%%%%%%%%%%%%%%%%%%%%%%%%%%%%%%%%%%%%%%%%
\begin{figure*}[ht]
\centering
 \includegraphics[width=.45\hsize]{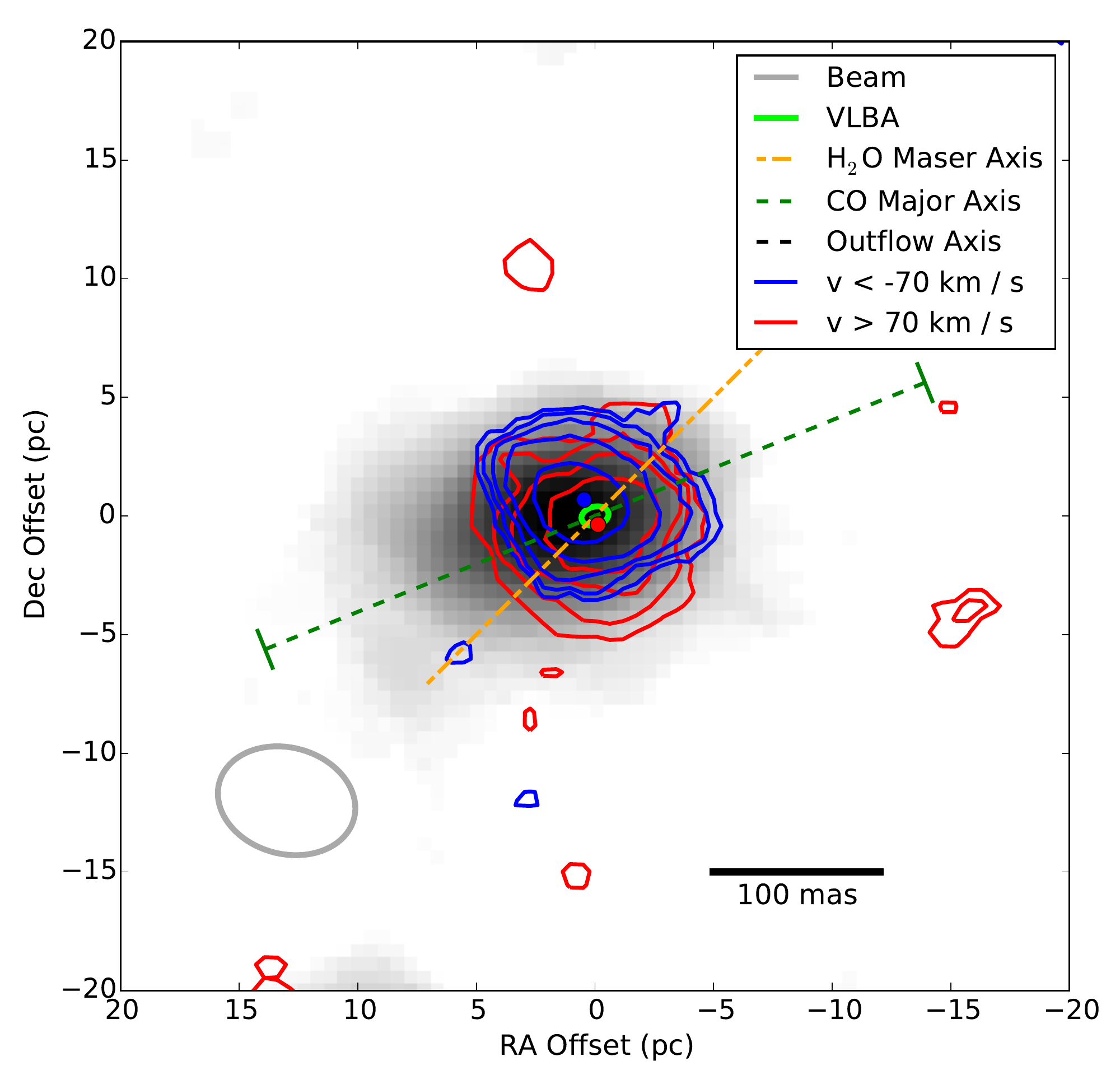}
 \includegraphics[width=.45\hsize]{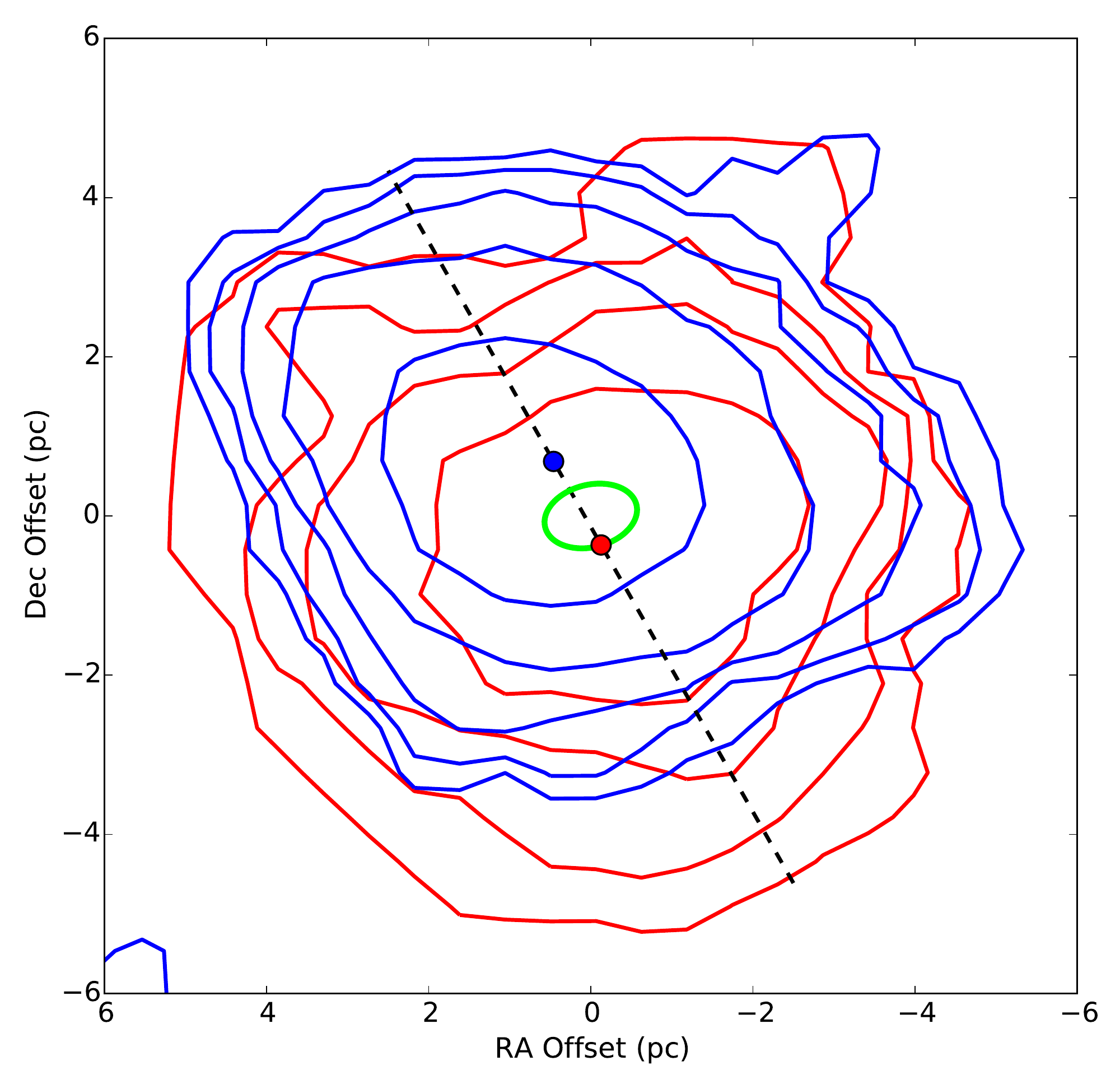}

\caption{{\em Left}: CO image integrated between radial velocities $-70$ and
$+70$~\kms\ relative to systemic, shown as grayscale linearly stretched between
0.30 and 7.0~Jy~\kms~beam\mone. Position offsets are presented in physical
units  and are measured with respect to the centroid of the 5~GHz VLBA image of
S1, the nuclear plasma disk whose orientation and location are illustrated by
the green ellipse. The 100~mas scale bar in the lower right corner indicates
the angular scale. The dashed green line (PA~112\deg) illustrates the CO image
major axis and the extraction axis for the position-velocity diagram (see text
for details). The end-caps on the major axis line show the three-pixel (24~mas,
or 1.8~pc) extraction width for the position-velocity diagram. Red and blue
contours trace the integrated CO emission over velocities $> +70$~\kms\ and $<
-70$~\kms, respectively. The contour levels are 0.30, 0.50, 0.83, 1.4, and
2.3~Jy~\kms~beam\mone. The alternating dashed orange line (PA~135\deg) shows
the orientation of the H$_2$O maser disk, stretched beyond its \about 1.7 pc
major axis to emphasize its orientation at the scale of the plot. {\em Right}:
Zoom-in on the high-velocity contours; the contour peaks are marked by colored
dots. The red and blue contours are systematically displaced from each other,
with their peaks separated by 1.2~pc along the axis PA~33\deg, which is traced
by the dashed black line. This displacement is indicative of a bipolar outflow
roughly orthogonal to the nuclear disk.
}
  \label{fig:comap}
\end{figure*}
%%%%%%%%%%%%%%%%%%%%%%%%%%%%%%%%%%%%%%%%%%%%%%%%%%%%%%%%%%%%%%%%%%%%%%%%%%%%

Beginning with the VLTI detection of the nucleus in NGC~1068 \citep{Jaffe04},
infrared (IR) observations have established the compact dimensions of AGN dusty
tori \citep{Elitzur08, AGN2}. Optical--near-IR reverberation lags in Type 1
Seyferts are consistent with sub-parsec torus inner radii \citep[e.g.,][and
references therein]{Vazquez15}. Further confirmation comes from recent ALMA
observations by \citeauthor{García-Burillo16} (2016; GB16 hereafter). They
detected molecular emission from the NGC~1068 torus, contained within \about 3
pc from the center. However, although the torus's compact dimensions are now
firmly established, its kinematics and dynamic origin remain uncertain. Water
maser observations provide support for the disk-wind scenario in Circinus
\citep{Greenhill03} and NGC~3079 \citep{Kondratko05} but the tight constraints
on strong maser emission do not allow a complete sampling of the molecular gas.

Here we report our own ALMA observations of thermal CO emission from the
NGC~1068 torus. We adopt a distance to the source of 14.4~Mpc
\citep{Bland-Hawthorn97}. The scale is $1\arcsec = 70$~pc.

\section{Observations and Data Processing}

We observed NGC~1068 in Band~9 ($\lambda \sim 450$\micron) during ALMA Cycle 2
(project code 2013.1.00014.S). Total time on source was 32.1~minutes. The
baseline range was 15--1466~m, and the synthetic restoring beam (angular
resolution) is $0\farcs084 \times 0\farcs064$, PA~$75\deg$, or $5.9 \times
4.5$~pc. Resulting images are insensitive to structures larger than about
270~pc (3.8\arcsec).  Source observations were interleaved with short scans of
the nearby bright calibrator source J0217+0144, located 6.5\deg\ from NGC~1068.
Spectral windows were tuned to observe three independent continuum bands, 2~GHz
bandwidth each, and one line window for CO ($J=6\rightarrow 5$; $\nu_0 =
691.47308$~GHz; hereafter, CO) in $\sim 0.21$~\kms\ channels spanning $\sim \pm
400$~\kms\ total bandwidth. Calibration and data reduction followed standard
procedures for ALMA observations in CASA (CASA v.4.3.1;
\citealt{2007ASPC..376..127M}). NGC~1068 proved bright enough for
self-calibration over scan intervals. Two rounds of phase-only self-calibration
reduced sidelobe artifacts and improved the background rms.

The photometric accuracy, \about 10\%, is limited by the photometric
uncertainties of the flux calibrators (3C454.3 and J2253+1608). The absolute
astrometric accuracy of these ALMA observations are limited by the transfer of
calibration solutions from the complex gain calibrators to NGC~1068. We
estimate that the final absolute astrometric accuracy\footnote{Based on
information provided by the
\href{https://help.almascience.org/index.php?/Knowledgebase/Article/View/319/6/what-is-the-astrometric-accuracy-of-alma}{ALMA
Helpdesk.}} of our observations is 1.3~pc (19~mas). This systematic uncertainty
is large compared to the \about 4~pc scale of the nuclear environment. We
therefore adopted an astrometric correction by aligning the Band~9 continuum
peak to the position of the active nucleus (cf.\ GB16).  In angular units, the
shift moves the Band~9 continuum peak 9.5~mas west and 11~mas south. In
physical units, the shift is 1.0~pc, which lies within the 1.3~pc systematic
uncertainty.

Images and data cubes were produced using Briggs weighting \citep{Briggs1995}
and multiscale CLEAN deconvolution \citep{2008ISTSP...2..793C}. The CO spectral
line cube was binned to 10~\kms\ channel-widths during imaging and
deconvolution. The background rms noise levels are 0.49~mJy~beam$^{-1}$ and
5.3~mJy~beam$^{-1}$ for Band~9 continuum and the CO channel maps, respectively.

%%%%%%%%%%%%%%%%%%%%%%%%%%%%%%%%%%%%%%%%%%%%%%%%%%%%%%%%%%%%%%%%%%%%%%%%%%%%
\begin{figure*}[t]
\centering
 \includegraphics[width=\textwidth]{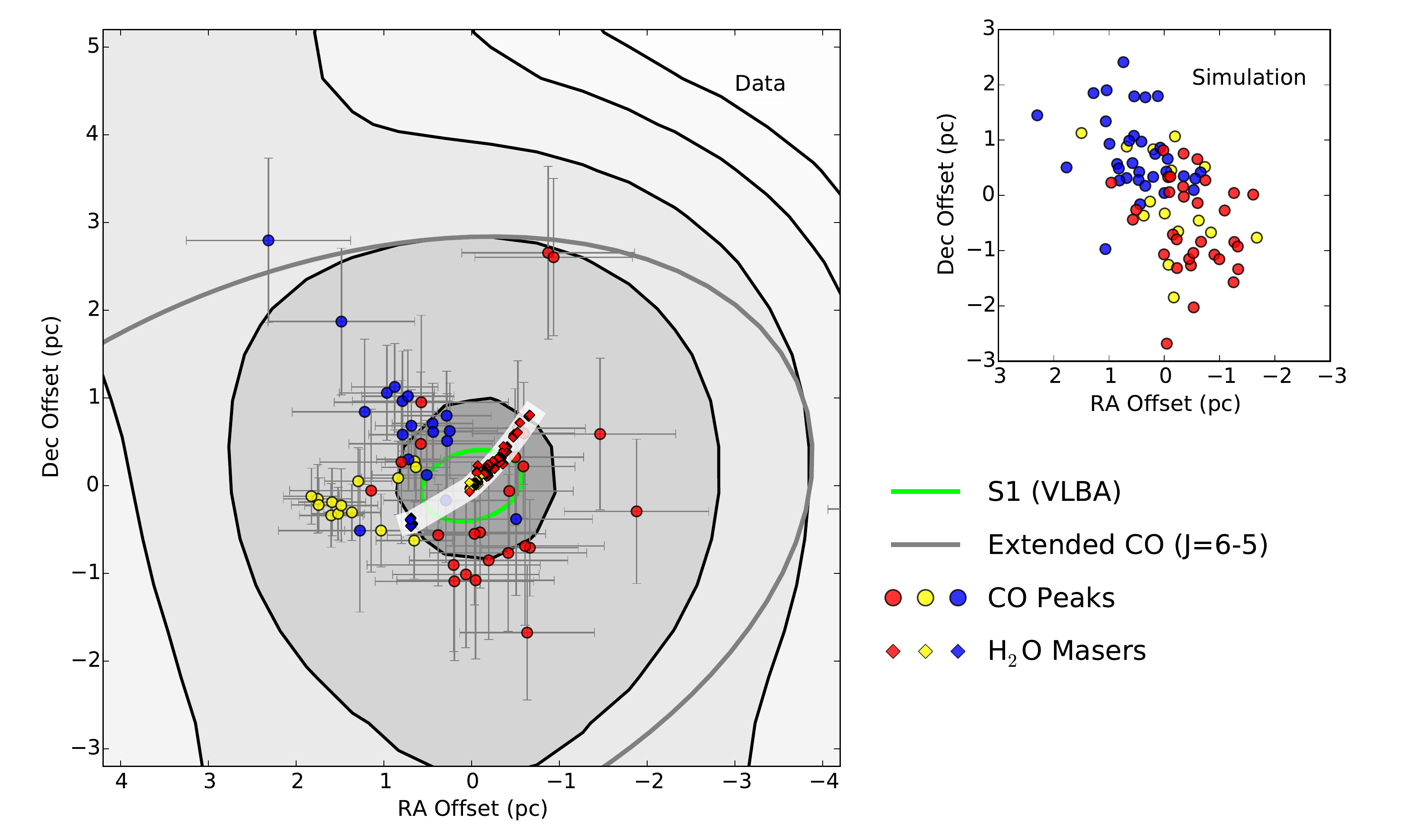}

\caption{{\em Left}: The locations of CO emission peaks, marked by colored
circles. The colors are coded by radial velocity bins relative to systemic: $v
> 70$~\kms\ (red), $-70 \leq v \leq 70$~\kms\ (yellow), and $v < -70$~\kms\
(blue). Following the same velocity color coding, the positions of H$_2$O maser
spots \citep{Greenhill97} are plotted as diamonds within the central, white
arcuate band. The location and scale of the VLBA radio continuum source is
indicated by the green ellipse. The outer gray ellipse illustrates the
orientation of the resolved, low velocity CO emission (see Fig.\
\ref{fig:comap}). Grayscale contours are from MERLIN 5~GHz observations of the
radio jet \citep{Gallimore04}; the MERLIN beam is 4.2~pc FWHM. {\em Upper
Right}: Spectral map from a simulation of a bipolar outflow with axis tilted
5\deg\ from the plane of the sky and rotated to PA~33\deg\ (see text for
details). Colors are as for the left panel.}
  \label{fig:spectroAstrometry}
\end{figure*}
%%%%%%%%%%%%%%%%%%%%%%%%%%%%%%%%%%%%%%%%%%%%%%%%%%%%%%%%%%%%%%%%%%%%%%%%%%%%

\section{Results}

GB16 performed similar observations of NGC~1068, and their continuum and
integrated line maps resemble ours. Like GB16, we detect continuum and
molecular emission located at the position of the nuclear radio source, S1, and
its associated H$_2$O vapor maser disk \citep{Greenhill97, Gallimore04}. The
detected CO spans the full $\pm$400 \kms\ spectral range.
Figure~\ref{fig:comap} shows the CO image of S1 integrated over three bins:
blueshifted, $v < -70$~\kms (blue contours); systemic, $-70 \leq v \leq
+70$~\kms (greyscale); and redshifted, $v > +70$~\kms (red contours), where $v$
is the observed radial velocity relative to systemic. The choice of $\pm
70$~\kms\ for velocity bin edges was based on initial inspection of the data
and the analysis of near-infrared H$_2$ emission \citep{Galliano02,
2009ApJ...691..749M}. At low $v$, the source resolves along PA~112\deg\ (green
dashed line), extending \about 10~pc southeast of the VLBA source.

The CO source is unresolved in the integrated blueshifted and redshifted
velocity bins. However, the blue contours are systematically displaced with
respect to the red ones along the dashed black axis in Figure~\ref{fig:comap};
in particular, the position of the blueshifted velocity peak is displaced
1.2~pc northeast (PA$\simeq 33\deg$) of the redshifted peak. The displacement
axis is roughly orthogonal ($78\deg\pm2\deg$) to the line of masers, which
trace a nearly edge-on disk. Thus \emph{the high-velocity CO emission is
consistent with a bipolar outflow along the AGN axis}, similar to the outflow
inferred from observations of ionized gas in the narrow line region of the
NGC~1068 nucleus \citep{Crenshaw00}.

Additional evidence for the CO bipolar outflow comes from the spectral map in
the left panel of Figure \ref{fig:spectroAstrometry}. Here we show the
locations of CO emission peaks from the full spectral line cube. We performed
spectroastrometry using find\_peaks from the photutils package of Astropy
\citep{Astropy13}. Peaks $ > 3.0\sigma$ on individual channel maps were
recorded, and their positions were refined using a local 2-D Gaussian fit. The
map shows that each high-velocity bin predominates in one of the two
hemispheres defined by the maser disk, as expected from a bipolar outflow
roughly aligned with the inner part of the radio jet, whose contours are also
shown. We propose that the CO emission arises from a bipolar outflow around the
axis of the AGN accretion disk. Molecular gas moving in the plane of the sky
contributes to the systemic-velocity bin, whereas gas on the near (far) side of
the plane appears in the blue (red) bin.

The outflow axis of ionized gas traced by optical emission lines in the
narrow-line region is inclined 5\deg\ out of the plane of the sky
\citep{Crenshaw00}, and we assume the same for the CO outflow. Can such a small
inclination produce the red-blue spatial separation observed in CO? To test
this possibility we conducted a numerical simulation of a bipolar outflow. We
employ a truncated cone geometry with a disk base diameter of 1~pc (comparable
to the H$_2$O maser disk), 4.4~pc total bicone height (based on the spread of
CO features in Figure~\ref{fig:spectroAstrometry}), and half-opening angle of
30\deg. To model the outflow, we randomly placed a total of $10^4$ clumps
inside the cone. Clumps were evenly distributed in cylindrical radius $\rho$
(i.e., radius projected onto the disk plane) and distance along the polar axis.
The direction of motion for a clump is vertical for clumps directly above and
below the disk (i.e., clumps with $\rho \leq 0.5$~pc) and radial from the edge
of the disk for clumps with $\rho > 0.5$~pc. We further assume uniform
acceleration from 30~\kms\ at the molecular disk to 800~\kms\ at the cone
height; these values were selected from interactive trials to give a reasonable
match to the observations. Finally, all clumps are equally luminous point
sources of CO emission, added as beam-shaped, two-dimensional Gaussians to the
planes of a simulated data cube. The cone was then inclined by 5\deg, and we
repeated the cube analyses on the simulated data. To simulate the effects of
measurement uncertainty, each measured peak position was displaced randomly,
assuming 0.5~pc errors, comparable to the relative, observed position
uncertainties. The simulation results shown in the right panel of Figure
\ref{fig:spectroAstrometry} broadly reproduce a parsec-scale asymmetry of
redshifted and blueshifted peaks similar to that observed (left panel). This
simple simulation was not meant to replicate any physical model or explain
every aspect of the data; in particular, the systemic-velocity emission is
concentrated to the east, at odds with the random distribution we use. Our
simulation is only meant to demonstrate that the small 5\deg\ tilt detected in
narrow line observations is sufficient to produce the red-blue asymmetry of the
excited CO emission.

%%%%%%%%%%%%%%%%%%%%%%%%%%%%%%%%%%%%%%%%%%%%%%%%%%%%%%%%%%%%%%%%%%%%%%%%%%%%
\begin{figure}[t]
\centering
 \includegraphics[width=\hsize]{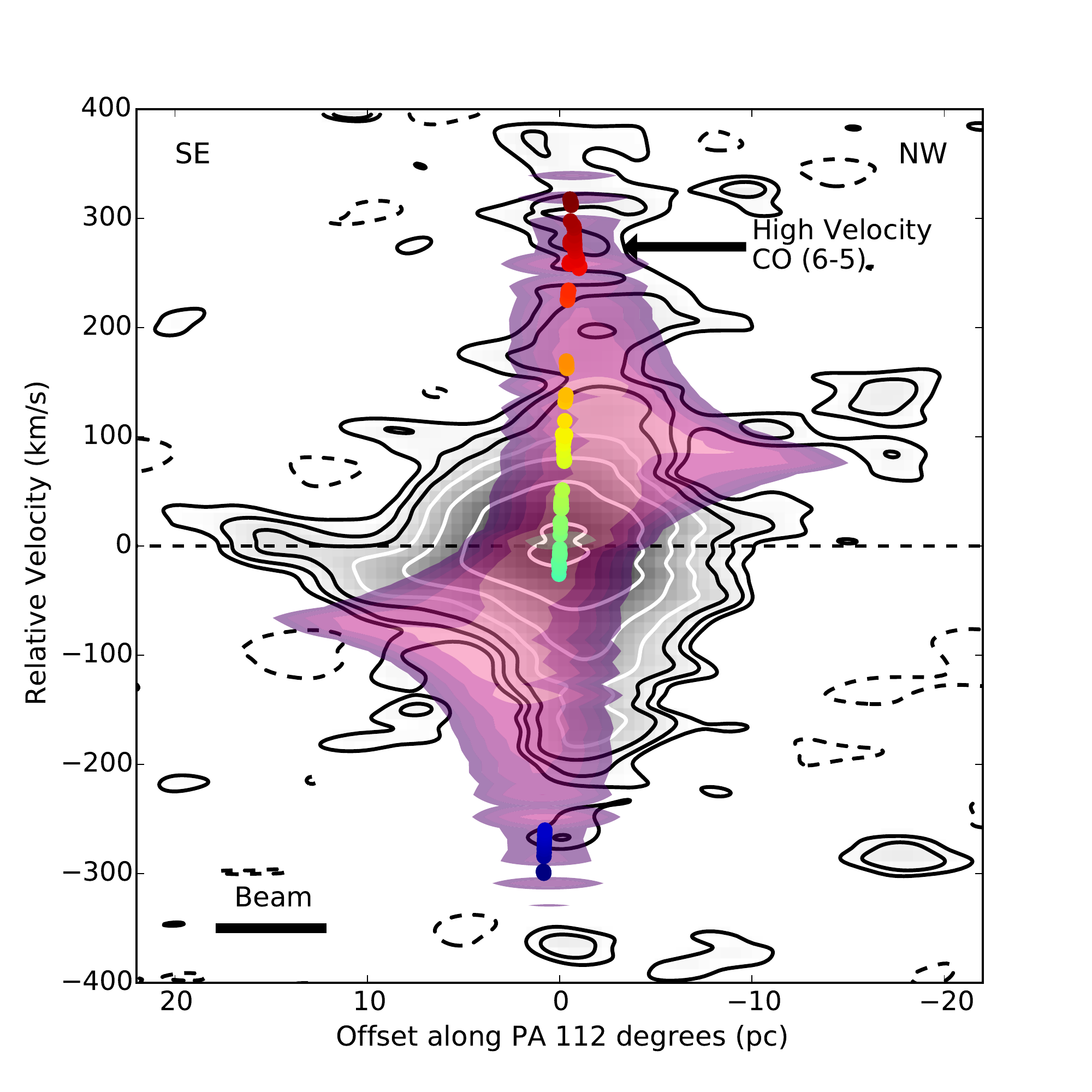}

\caption{CO position-velocity diagram, with position varying along the CO major
axis (the green dashed line at PA 112\deg\ in Figure~\ref{fig:comap}). The data
are shown as grayscale with contours at -4.7, 4.7, 7.2, 11.0, 16.8, 25.7, 39.2,
and 60 mJy~beam$^{-1}$. The beamwidth in the direction of the position slice is
shown at the lower left. The rainbow-colored circles mark the \pv\ measurements
of the H$_2$O masers. The transparent violet shading traces the emission
expected from Keplerian rotation around a 1.5\x\E7\,\Mo\ central mass (see text
for details).
}
  \label{fig:pv}
\end{figure}
%%%%%%%%%%%%%%%%%%%%%%%%%%%%%%%%%%%%%%%%%%%%%%%%%%%%%%%%%%%%%%%%%%%%%%%%%%%%

In addition to the outflow component described above, we also find evidence for
rotation consistent with the compact H$_2$O maser disk. Figure~\ref{fig:pv}
shows the position-velocity (\pv) diagram constructed by slicing the CO
spectral line cube along the axis tracing the extended systemic emission.
Surface brightnesses were averaged within the 3-pixel extraction width (green
dashed lines in Figure~\ref{fig:comap}). The data were smoothed using a
Savitzky-Golay filter (7 channels, polynomial order 3; \citealt{SavitzkyGolay})
along the velocity axis for display purposes. The diagram reveals CO emission
at very high velocities, with significant ($> 3\sigma$) detections approaching
$v = \pm 400$~\kms\ relative to systemic. Also shown are the H$_2$O maser
spots, appearing as a nearly vertical chain centered on the origin. Such a
straight line in the \pv\ diagram is the hallmark of a rotating ring whose
angular velocity is the line slope \citep[e.g.,][]{Pestalozzi09}. The fact that
the maser line is so straight and narrow implies that all maser spots are
located within a radially thin annulus at the disk inner radius. The ALMA data
show a similar elongated structure aligned with the maser chain, only covering
a wider spread of positions. This component can be interpreted as CO emission
from a wider range of rotating regions close to the disk innermost boundary.
The thermal CO emission traces these molecular regions more fully than the
masers, because strong maser emission is not produced without the fulfilment of
some strict constraints on the physical parameters. For example, these
constraints likely explain the dearth of blue-shifted maser spots.
Additionally, the CO data also show another elongated component, this one
closer to the horizontal axis. Its smaller slope implies a lower angular
velocity, which in turn implies a larger radius and can be interpreted as
emission from the outer regions of the CO distribution.

An elongated structure in the \pv\ diagram could also be produced by a
collimated jet, in which case the slope is directly related to the jet
inclination from the line of sight \citep{Pestalozzi09}. The existence of two
such features with different slopes and the overlap of one of them with the
maser line make a jet-based explanation unlikely, indicating that the CO
velocity field contains a kinematic component consistent with rotation.

To
explore this possibility we simulated the \pv\ diagram of an edge-on disk with
kinematics based on Keplerian rotation around a 1.5\x\E7\,\Mo\ central mass
that best fits the H$_2$O maser disk observations \citep{Lodato03} but with
line of nodes rotated 23\deg\ to match the \pv\ slice orientation. To highlight
the expected contribution of such a disk, its emissivity was scaled
proportional to $r^{-2}$, where $r$ is the disk radial coordinate. The
simulated data were beam-smoothed and binned into 10~\kms\ channels to compare
with the ALMA observations. The result is shown as transparent, violet-shaded
contours in Figure~\ref{fig:pv}; note the expected ``tilted bow-tie" shape
\citep[cf.][]{Pestalozzi09}.

This simulation is not intended to fit the CO data but to illustrate the
appearance of a rotating disk on the \pv\ diagram. The overall similarity with
the data suggests the presence of rotation on top of the radial outflow, where
the latter fills in the regions in the \pv\ diagram inaccessible to rotational
motion. The emerging picture is in agreement with the disk-wind model, first
put forward by \cite{Blandford_Payne}. Such a wind can be visualized as a stack
of rotating disks that are accelerated away from the base of the outflow. Each
disk in the stack could in turn be described by an analysis similar to the one
that produced the violet-shaded structure in Figure~\ref{fig:pv}. It thus seems
that the simplest interpretation is that the CO emission traces a combination
of rotation and radial outflow on pc~scales surrounding the central engine,
consistent with the disk-wind scenario. From the \cite{Kondratko05}
observations of H$_2$O masers in NGC~3079, \cite{EH09} suggest that the
disk-wind launch velocity is roughly 10\% of the local Keplerian velocity, or a
few $\times 10$~\kms\ in the case of NGC~1068. Subsequent acceleration can lead
to the observed radial velocities of order several 100~\kms\
\citep[e.g.,][]{Blandford_Payne, Emmering92, Kartje99}.

\section{Discussion}

% * The PA of the masers is about 135 degrees (using extreme position masers to
%   make a line)
% * The PA of the CO outflow is about 33 degrees (peak to peak on Figure 1)
% * Therefore, delta PA = 102 degrees (on the east side)
% * But really delta PA = 180 - 102 degrees = 78 degrees (minimum angle between
%   maser disk and CO outflow on the west side)
% * Uncertainty of delta PA ~ 2 degrees

The misalignment between the maser and radio disks has been a puzzle ever since
the \cite{Gallimore04} VLBA observations. Our ALMA data show that the systemic
velocity CO emission is closely aligned with the S1 radio disk, enhancing the
case for this to be the true orientation of the AGN accretion disk. The maser
inclination may be attributed to the foundations of the emission process:
strong maser amplification requires tight velocity coherence along the line of
sight. It is possible that some transient process created favorable conditions
for such coherence on the disk inner boundary at an angle to the disk plane.
Either direction is nearly orthogonal to the CO outflow axis, which is rotated
79\deg\ clockwise from the S1 axis and 78\deg\ counterclockwise from the maser
axis. Thus the NGC~1068 nucleus has a bipolar outflow, seen in the
high-velocity CO, radio jet, and narrow optical emission lines. The outflow is
roughly orthogonal to the accretion disk, which is traced by the systemic
velocity CO, radio continuum emission and H$_2$O masers.

Having detected a single CO transition, it is difficult to derive strong
constraints on the physical parameters of emitting clouds. Some indications can
still be obtained from the $J = 6 \to 5$ brightness temperature, which peaks at
31.1 K. Detailed radiative transfer calculations of CO line emission show that
such brightness temperature requires H$_2$ density in excess of \about \E5 \cc,
temperatures higher than \about 50 K and CO column densities $N_{\rm CO} \ga$
\E{17} cm$^{-2}$(\kms)$^{-1}$. From modeling of IR emission from AGN, torus
clouds are expected to have hydrogen column densities of $N_H$ \about \E{23}
\cs, assuming the standard dust abundance $N_H = 2\x\E{21}A_V$ \cs\
\citep{AGN2}. An intrinsic linewidth of \about 10~\kms\ would then imply CO
abundance of \about \E{-5}, similar to Galactic values.

Forthcoming publications will present detailed models and report on additional
molecular and IR continuum observations of NGC~1068. The CO data reported here
already provide the most compelling evidence yet in support of the disk-wind
scenario for the AGN obscuring torus.

\acknowledgments This paper makes use of the following ALMA data:
ADS/JAO.ALMA\#2013.1.00014.S. ALMA is a partnership of ESO (representing its
member states), NSF (USA) and NINS (Japan), together with NRC (Canada), NSC and
ASIAA (Taiwan), and KASI (Republic of Korea), in cooperation with the Republic
of Chile. The Joint ALMA Observatory is operated by ESO, AUI/NRAO and NAOJ. The
National Radio Astronomy Observatory is a facility of the National Science
Foundation operated under cooperative agreement by Associated Universities,
Inc. This research made use of Astropy, a community-developed core Python
package for Astronomy \citep{Astropy13}. RN acknowledges support by FONDECYT
grant No. 3140436. RM acknowledges support from the UK Science and Technology
Facilities Council (STFC).

\facility{ALMA} \software{CASA, astropy}

%\bibliographystyle{aasjournal}
%\bibliography{Letter1068-v09,AGN}
%
%\end{document}

\end{document}